\documentclass[aps,pre,twocolumn,superscriptaddress,tight]{revtex4-1}

\usepackage{amssymb}
\usepackage{amsfonts}
\usepackage{color}
\usepackage{graphicx}
\usepackage{amsmath}
\usepackage{times}
\usepackage{bm}
\usepackage{float}
\usepackage{lipsum}
\usepackage{import}
\usepackage{enumerate}
\usepackage[colorlinks,linkcolor=blue,anchorcolor=blue,citecolor=blue]{hyperref}

\begin{document}

\title{Sustaining the dynamics of Kuramoto model by adaptable reservoir computer}

\author{Haibo Luo}
\thanks{These authors contributed equally to this work}
\affiliation{School of Physics and Information Technology, Shaanxi Normal University, Xi'an 710062, China}

\author{Mengru Wang}
\thanks{These authors contributed equally to this work}
\affiliation{School of Physics and Information Technology, Shaanxi Normal University, Xi'an 710062, China}

\author{Yao Du}
\affiliation{School of Physics and Information Technology, Shaanxi Normal University, Xi'an 710062, China}

\author{Huawei Fan}
\affiliation{School of Science, Xi’an University of Posts and Telecommunications, Xi'an 710121, China}

\author{Yizhen Yu}
\affiliation{School of Physics and Information Technology, Shaanxi Normal University, Xi'an 710062, China}

\author{Xingang Wang}
\email{E-mail address: wangxg@snnu.edu.cn}
\affiliation{School of Physics and Information Technology, Shaanxi Normal University, Xi'an 710062, China}

\begin{abstract}
A scenario frequently encountered in real-world complex systems is the temporary failure of a few components. For systems whose functionality hinges on the collective dynamics of the interacting components, a viable approach to dealing with the failures is replacing the malfunctioning components with their backups, so that the collective dynamics of the systems can be precisely maintained for a duration long enough to resolve the problem. Here, taking the paradigmatic Kuramoto model as an example and considering the scenario of oscillator failures, we propose substituting the failed oscillators with digital twins trained by the data measured from the system evolution. Specifically, leveraging the technique of adaptable reservoir computer (RC) in machine learning, we demonstrate that a \emph{single, small-size} RC is able to substitute {\it any} oscillator in the Kuramoto model such that the time evolution of the synchronization order parameter of the repaired system is identical to that of the original system for a certain time period. The performance of adaptable RC is evaluated in various contexts, and it is found that the sustaining period is influenced by multiple factors, including the size of the training dataset, the overall coupling strength of the system, and the number of substituted oscillators. Additionally, though the synchronization order parameter diverges from the ground truth in the long-term running, the functional networks of the oscillators are still faithfully sustained by the machine substitutions.
\end{abstract}
\maketitle

\section{Introduction}

Data-based, model-free inference of chaotic dynamics by the technique of reservoir computer (RC) in machine learning has gained significant attention in recent years~\cite{RC:Maass2002,RC:Jaeger,RC:Pathak2017,RC:LZ2018,RC:Fan,RC:digtwin,RC:DY2024,RC:NC2024perspective,RC:adaptiveRCLYC}. From the point of view of dynamical systems, an RC can be regarded as a complex network composed of nonlinear elements which, driven by the input data, generates the output data through a readout function. Compared to other deep learning techniques in machine learning, RC contains only a single hidden layer, namely the reservoir, and, except for the output matrix that is estimated from the measured data through a training process, all other settings of the machine are fixed at the construction. The simple architecture renders RC an ideal candidate for applications in which training data are scarce and computational resources are limited. Though structurally simple, RC has shown great potential in many data-oriented applications, especially for temporal sequences~\cite{RC:lukosevicius2009,RC:Tanaka2019,RC:Book}. For instance, studies have shown that a properly trained RC is able to predict accurately the state evolution of typical chaotic systems for about a dozen Lyapunov times~\cite{RC:Jaeger,RC:Pathak2017,RC:LZ2018}, which is much longer than the prediction horizon of the traditional methods developed in nonlinear science. Besides predicting the short-term state evolution, RC is also able to replicate the long-term statistical properties of chaotic systems~\cite{RC:Pathak2017,RC:LZ2018}, namely the ``climate" of the system dynamics.

Whereas early studies of RC have been focused on the inference of a single chaotic system which generates both the training and testing data, recent research has begun to explore the capability of RC for inferring the dynamics of multiple chaotic systems and dynamics that are not included in the training data~\cite{RC:CK2020,RC:Guo2021,KLW:2021,RC:Kim2021,RC:FHW2021,RC:XR2021,RC:ZH2021,HWFan2022,RC:multistability2022,PARC:Jalan2024,RC:LHB2024}. In particular, introducing a parameter-control channel into the standard architecture of RC, a new scheme of RC, namely the scheme of parameter-aware RC~\cite{KLW:2021} or adaptable RC~\cite{RC:adaptiveRCLYC}, has been proposed for inferring new dynamics not included in the training data~\cite{KLW:2021,RC:Kim2021,RC:FHW2021,RC:XR2021,RC:ZH2021,HWFan2022,RC:multistability2022,PARC:Jalan2024,RC:LHB2024}. In this new scheme, the machine is trained by only the time series of several states of a dynamical system, but it is able to infer the ``dynamics climate” of new states not seen in the training data. The significance of adaptable RC is reflected in the capability of knowledge transfer from the training data to new systems, which has important implications for the inference of dynamics transitions in complex dynamical systems, saying, for example, predicting the tipping points of system collapses~\cite{KLW:2021}, anticipating the critical coupling required for synchronization in coupled oscillators~\cite{RC:FHW2021}, and inferring the bifurcation diagram of chaotic circuits~\cite{RC:LHB2024}. More recently, this technique has been exploited for storing and retrieving multiple chaotic attractors governed by entirely different dynamics, namely the design of multifunctional RC~\cite{KLW2024,YD:MFRC}.

In addition to low-dimensional chaotic systems, RC has also been exploited for predicting spatially extended systems in recent studies. Compared to low-dimensional chaos, the inference of spatiotemporal dynamics is more challenging and demands vast training data and computational resources, known as the ``curse of dimensionality”~\cite{COD}. One approach to coping with the challenge is adopting the scheme of parallel RC~\cite{RC:digtwin,RC:Pathak2018,RC:Parlitz2018,RC:ParallelPRL2022,RC:AtmosphereForecastOtt2020}, in which the high-dimensional system is decomposed into an ensemble of low-dimensional, local elements, with each element being emulated by a unique RC. As each RC mimics only the dynamics of a specific element, the machines are of small size and can be easily trained. To infer the dynamics of the high-dimensional system, the machines are coupled according to the interacting relationship of the local elements and updated simultaneously. The scheme of parallel RC has been applied successfully to the inference of a variety of spatially extended systems, e.g., the spatiotemporal chaos described by the Kuramoto-Sivashinsky equation~\cite{RC:Pathak2018}, the spiral wave patterns in two-dimensional excitable media~\cite{RC:Parlitz2018}, the collective dynamics of coupled oscillators on regular networks~\cite{RC:digtwin}, and even the atmospheric dynamics of the entire globe~\cite{RC:AtmosphereForecastOtt2020}. In these studies, a common feature is that the systems are homogeneous, i.e., the local elements are of identical dynamics and are coupled on regular networks. As many real-world systems are represented by complex networks of non-identical dynamical elements, the generalization of the parallel RC scheme to complex systems of heterogeneous dynamics is thus of great significance and attracts substantial interest in the fields of complex systems and machine learning. In Ref.~\cite{RC:ParallelPRL2022}, the authors exploited the parallel scheme to forecast the dynamics of homogeneous complex networks (all nodes are of the same number of connections) composed of non-identical phase oscillators. It is shown that, compared to the nonparallel scheme, the parallel scheme not only leads to a significant reduction in computational cost, but also extends the prediction horizon by approximately an order of magnitude. It is worth noting that in Ref.~\cite{RC:ParallelPRL2022}, the number of RCs is the same as the system size, and each oscillator is emulated by a unique RC that is trained by the time series of both the target oscillator and its neighbors.

Inspired by the studies in Ref.~\cite{RC:ParallelPRL2022}, we attempt to address the inference of spatiotemporal dynamics of heterogeneous complex systems from the perspective of element substitution by asking the following question: {\it Is it possible to train a single, small-size RC by the time series of the system elements such that the same machine can substitute any element while ensuring that the collective dynamics of the system remain unchanged over a sufficiently long period?} While our overarching goal is to resolve this question for general complex systems of both heterogeneous structures and dynamics, here we make the first step towards this goal by considering the dynamics of an ensemble of globally coupled non-identical phase oscillators, namely the classical Kuramoto model. Our main finding in the present work is that, by generalizing the technique of adaptable RC proposed recently in machine learning, the objective is accomplishable for the classical Kuramoto model. We shall present in the following section the classical Kuramoto model and the target state to be sustained by the machine. The technique of generalized adaptable RC will be introduced in Sec. III. The performance of adaptive RC in sustaining the system dynamics by substituting a randomly chosen oscillator will be reported in Sec. IV, together with a detailed analysis of the impacts of the dataset and system parameters on the machine performance. Finally, discussions and conclusions will be given in Sec. V.

\section{Model}

The classical Kuramoto model adopted in our studies is described by the set of equations~\cite{Kuramoto:Book}
\begin{equation}\label{model}
\dot{\theta_i}=\omega_i+\frac{K}{N}\sum_{j=1}^N\sin(\theta_j-\theta_i),
\end{equation}
where $i,j=1,2,\ldots,N$ denotes the oscillator indices, $\theta_i(t)$ represents the instant phase of oscillator $i$, $\omega_i$ is the natural frequency of oscillator $i$, and $K$ stands for the uniform coupling strength. The natural frequencies are drawn randomly from the range $(-\pi/2, \pi/2)$, and the initial phases of the oscillators are randomly chosen from the range $(0, 2\pi]$. We characterize the collective dynamics of the system by the synchronization order parameter
\begin{equation}\label{orderparameter}
R(t)=\frac{1}{N}\left| \sum_{j=1}^N e^{\mathbf{i}\theta_j(t)} \right|,
\end{equation}
with $\mathbf{i}=\sqrt{-1}$ being the imaginary unit and $|\cdot|$ representing the modulus function. We have $R(t) \in [0, 1]$, with $R(t) = 0$ and $1$ corresponding to the completely desynchronized and globally synchronized states, respectively. In our studies, we set the system size as $N = 130$, and Eq.~(\ref{model}) is solved numerically by the fourth-order Runge–Kutta algorithm with the time step $\delta t = 0.01$.

\begin{figure}[tbp]
\begin{center}
\includegraphics[width=0.8\linewidth]{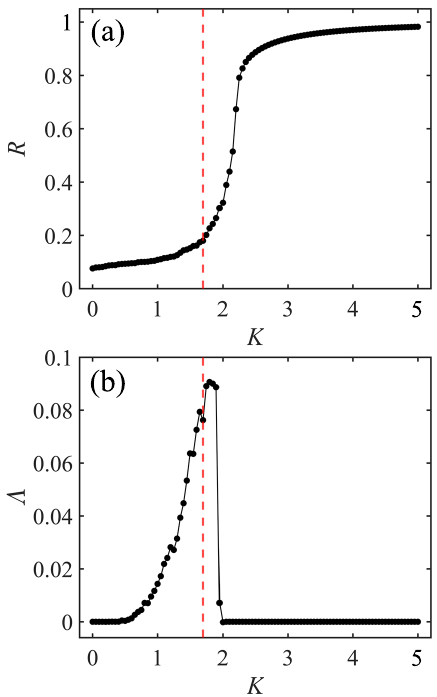}
\centering\caption{For the classical Kuraoto model containing $N = 130$ non-identical phase oscillators, the variation of the time-averaged synchronization order parameter, $R$, with respect to the uniform coupling strength, $K$ in (a) and the variation of the largest Lyapunov exponent, $\Lambda$, with respect to $K$ in (b). Red dashed lines denote the sampling state chosen for testifying the performance of the machine}
\vspace{-0.3cm}
\label{fig1}
\end{center}
\end{figure}

We demonstrate first the transition of the system dynamics with respect to the coupling strength. Shown in Fig.~\ref{fig1}(a) is the variation of the time-averaged synchronization order parameter, $R$, with respect to the uniform coupling strength, $K$. In calculating $R$ for each coupling strength, a transient period of $T = 500$ is first discarded to avoid the impact of the initial conditions, and then $R$ is obtained by averaging $R(t)$ over a period of $T = 2 \times 10^3$. We see in Fig.~\ref{fig1}(a) that $R$ stays around $0.1$for $K<K_c \approx 1$ and, after a quick increase in the transition regime $K \in (K_c, 2.5)$, it reaches a large value about $0.9$. After that, $R$ approaches $1$ gradually as $K$ increases. To characterize the chaotic nature of the system dynamics in the process of synchronization transition, we plot in Fig.~\ref{fig1}(b) the variation of the largest Lyapunov exponent, $\Lambda$, with respect to $K$. Still, the value of $\Lambda$ is averaged over a period of $T =2\times 10^3$. We see that $\Lambda$ is positive when $K \in (1, 2)$, indicating that the system presents chaotic motions in this interval.

The sampling state we adopt to illustrate the performance of the new machine-learning technique is chosen at $K = 1.7$, by which the time-averaged synchronization order parameter is about $R = 0.18$ and the largest Lyapunov exponent of the system dynamics is about $\Lambda= 0.075$. Given that the time series of all oscillators in the past are available and the natural frequencies of the oscillators are also known {\it a priori}, our objective is to train a single, small-size RC capable of substituting any phase oscillator in the system while maintaining the time evolution of the synchronization order parameter unchanged over a sufficiently long period.

\section{Method}

The technique of adaptable RC adopted in our studies is generalized from the one proposed in Refs.~\cite{RC:adaptiveRCLYC,KLW:2021,RC:FHW2021} by incorporating coupling signals into the input channel. In specific, the machine is composed of three modules: the input layer, the hidden layer (reservoir network), and the output layer. The input layer is characterized by the matrix $\bm{W}_{in} \in \mathbb{R}^{D_r \times D_{in}}$, which couples the input vector $\tilde{\bm{u}}(t)\in \mathbb{R}^{D_{in}}$ into the reservoir network. The input vector is expressed as $\tilde{\bm{u}}_{\beta}(t) = [\bm{u}^s_{\beta} (t), \bm{u}^c_{\beta} (t), \beta(t)]^T$. Here, $\bm{u}^s_{\beta} (t)\in \mathbb{R}^{D_s}$ denotes the state vector of the target oscillator to be substituted by the machine, which is inputted into the machine through the state channel; $\bm{u}^c_{\beta} (t)$ represents the coupling vector the target oscillator receives from other oscillators in the system, which is inputted into the machine through the coupling channel; $\beta(t)$ is the intrinsic parameter of the target oscillator, which is inputted through the parameter control channel. The elements of the input matrix, $\mathbf{W}_{in}$, are randomly drawn from a uniform distribution within the range $[-\sigma, \sigma]$. The reservoir network contains $D_r$ nodes, with the initial states of the nodes being randomly chosen from the interval $[-1, 1]$. The states of the nodes in the reservoir network, $\bm{r}(t) \in \mathbb{R}^{D_r}$, are updated as
\begin{equation}\label{rc1}
\bm{r}(t+\Delta t)=(1-\alpha)\bm{r}(t)+\alpha\tanh[\bm {A}\bm{r}(t)+\bm{W}_{in}\tilde{\bm{u}}(t)].
\end{equation}
Here, $\Delta t$ is the time step used for updating the reservoir network (for the sake of simplicity, we set it as the time interval of data sampling), $\alpha\in (0, 1]$ is the leaking rate, $\tanh$ represents the hyperbolic tangent function, $\bm{A} \in \mathbb{R}^{D_r \times D_r}$ is a weighted adjacency matrix representing the coupling relationship between nodes in the reservoir network. The adjacency matrix $\bm{A}$ is constructed as a sparse random Erd\"{o}s-R\'{e}nyi matrix: with the probability $p$, each element of the matrix is arranged a nonzero value drawn randomly from the interval $[-1,1]$. The matrix $\bm{A}$ is rescaled to make its spectral radius equal $\lambda$. The output layer is characterized by the matrix $\bm{W}_{out}\in \mathbb{R}^{D_{out} \times D_r}$, which generates the output vector, $\bm{v}(t) \in \mathbb{R}^{D_{out}}$, by the operation
\begin{equation}\label{rc2}
\bm{v}(t+\Delta t)=\bm{W}_{out}\bm{\tilde{r}}(t+\Delta t),
\end{equation}
The output matrix, $\bm{W}_{out}$, is to be estimated from the measured data through a training process. Except $\bm{W}_{out}$, all other parameters of the RC, e.g., $\alpha$, $\bm{W}_{in}$ and $\bm{A}$, are fixed at the construction.

The implementation of the machine contains three phases: training, validation, and application. The mission of the training phase is to find a suitable output matrix, $\bm{W}_{out}$, so that the output vector $\bm{v}(t+\Delta t)$ as calculated by Eq.~(\ref{rc2}) is as close as possible to the state vector $\bm{u}^s_{\beta}(t + \Delta t)$ for $t = \Delta t,...,L\Delta t$, with $L$ the length of the training series. This can be achieved by minimizing the cost function with respect to $\bm{W}_{out}$, which gives~\cite{RC:Pathak2017,RC:LZ2018}
\begin{equation}\label{rc3}
\bm{W}_{out}=\bm{U}\bm{V}^T(\bm{V}\bm{V}^T+\eta\mathbb{I})^{-1},
\end{equation}
where $\bm{V}\in \mathbb{R}^{D_r\times L}$ is the state matrix whose $k$th column is $\bm{r}(k\Delta t)$, $\bm{U}\in \mathbb{R}^{D_{out}\times L}$ is a matrix whose $k$th column is $\bm{u}^s_{\beta}(k\Delta t)$, $\mathbb{I}$ is the identity matrix, and $\eta$ is the ridge regression parameter for avoiding the overfitting. Please note that for the task of oscillator substitution, the state vector, $\bm{u}^s$, and the output vector, $\bm{v}$, are of the same dimension, i.e. $D_{out} = D_s$.

The machine that performs well on the training data might not perform equally well on the testing data. The finding of the optimal machine performing well on both the training and testing data is the mission for the validating phase. The set of hyperparameters to be optimized in the machine include $D_r$ (the size of the reservoir network), $p$ (the density of the adjacency matrix $\bm{A}$), $\sigma$ (the range defining the elements of the input matrix), $\lambda$ (the spectral radius of the adjacency matrix $\bm{A}$), $\eta$ (the regression coefficient), and $\alpha$ (the leaking rate). In our studies, the optimal hyperparameters are obtained by scanning each hyperparameter over a certain range in the parameter space using conventional optimization algorithms such as the Bayesian and surrogate optimization algorithms~\cite{KLW:2021}. 

Having obtained the optimal machine, we then utilize it to substitute an oscillator randomly chosen in the system, namely the application phase. In doing this, we replace $\bm{u}^s_{\beta}(t)$ with $\bm{v}(t)$ (so that the machine is operating in the closed-loop mode), while setting $\beta$ as the intrinsic parameter of the substituted oscillator. The machine is coupled with other oscillators in the system as the substituted oscillator. Specifically, the machine receives coupling signals, $\bm{u}^c_{\beta}(t)$, from other oscillators in exactly the same way as the oscillator that is substituted by the machine and, in the meantime, the output vector, $\bm{v}(t)$, is also coupled to the same set of neighbors as the substituted oscillator in the system. We note that while the coupling relationship of the system elements is known {\it a priori} (the oscillators are globally coupled in the classical Kuramoto model), the coupling functions and the oscillator dynamics are unknown.

\section{Results}

\subsection{Datasets and machine settings}

We start by preparing the datasets used in machine training and optimization. The instant phase of the $i$th oscillator is represented by the vector $\bm{u}^s_i (t) =[\sin\theta_i (t), \cos \theta_i (t)]^T$ , and data points are sampled by the time interval $\tau = 5$ (i.e. $\Delta t = 0.05$). The time series are recorded for all oscillators, and each time series contains $\hat{L} = 8000$ data points. We split the time series into two segments. The first segment contains $L = 6000$ data points, which are used for training the output matrix. The second segment contains $L'= 2000$ data points, which are used for optimizing the machine hyperparameters. The training (validating) dataset is the concatenation of $m$ time-series stacks, $\bm{U} = ({\tilde{\bm{u}}_1, \tilde{\bm{u}}_2, \ldots , \tilde{\bm{u}}_m})$, with $\tilde{\bm{u}}_i \in \mathbb{R}^{D_{in}\times L}$ ($\mathbb{R}^{D_{in}\times L'}$) containing $L$ ($L'$)  points. Data point in the $i$th stack is represented by the vector $\tilde{\bm{u}}_i(t) = [\bm{u}^s_i(t), \bm{u}^c_i(t), \omega_i]^T$, which has the dimension $D_{in} = 2N + 1$. Here, $\bm{u}^s_i(t) = \bm{u}_i(t)$ stands for the state vector of the $i$th oscillator, which is inputted into the reservoir through the state channel; $\bm{u}^c_i = (\bm{u}_1,\ldots,\bm{u}_{i-1},\bm{u}_{i+1},\ldots,\bm{u}_N)^T$ denotes the coupling signals that oscillator $i$ receives from other oscillators in the system, which is inputted through the coupling channel; $\omega_i$ characterizes the natural frequency of oscillator $i$, which is inputted through the parameter-control channel. The oscillators ($m$ in total) generating the datasets are defined as the sampling oscillators, which are chosen randomly in the system. As an illustration, we set the number of sampled oscillators as $m = 20$, which is about $15\%$ of the system size. The length of the training and validating series thus are $m\times L = 1.2\times 10^5$ and $m \times L'= 4 \times 10^4$, respectively. We note that the time series of each oscillator is used $m$ times in constructing the datasets, and the time series of the control parameter, $\omega(t)$, is a step-function [i.e., $\omega(t) \equiv \omega_i$ for $\tilde{\bm{u}}_i$].

The processes of machine training and optimization are the same as the standard technique of parameter-aware RC~\cite{KLW:2021,RC:Kim2021,RC:FHW2021,RC:XR2021,RC:ZH2021}, except that the input data also includes the coupling signals, $\bm{u}^c_i (t) \in \mathbb{R}^{2(N -1)}$. Specifically, in the training phase, the input vector is $\tilde{\bm{u}}_i(t)$, and the cost function for estimating the output matrix is defined as the Euclidean distance ($L^2$ norm) between the output vector $\bm{v}(t + \Delta t)$ and the true state vector $\bm{u}^s_i (t+\Delta t)$. In this phase, the machine is operating in the open-loop mode. In the validation phase, we fix the reservoir size as $D_r = 1000$ for simplicity, while optimizing the other machine hyperparameters by minimizing the cost function over the validating dataset. The ranges over which the hyperparameters are tuned are $[0, 1]$ for $p$ (the density of the reservoir network), $[0, 3]$ for $\sigma$ (the range of the input matrix), $[0, 3]$ for $\lambda$ (the spectral radius of the reservoir), [$1 \times 10^{-10} , 1 \times 10^{-2} ]$ for $\eta$ (the regression coefficient), and $[0, 1]$ for $\alpha$ (the leaking rate). In this phase, the machine is operating in the closed-loop mode by replacing $\bm{u}^s_i (t)$ with $\bm{v}(t)$ in the state channel (the inputs of the coupling and parameter-control channels are still from the validating data.) The optimal hyperparameters are obtained after $300$ trials searching in the parameter space with the help of the ``optimoptions” function in MATLAB. The validation phase ends up with the set of optimal hyperparameters $(p, \lambda, \sigma, \alpha, \eta) = (0.406, 0.2566, 0.9791, 0.6687, 7 \times 10^{-3})$, which, together with the associated output matrix, defines the optimal machine to be used in the application phase.

In the application phase, the machine is first running in the open-loop mode by inputting a ``warm-up” series (containing $\tau= 100$ data points), and then operating in the closed-loop mode when it is used to substitute the target oscillator (which may or may not be included the sampling set). The ``warm-up” series, $\tilde{\bm{u}}_i(t)$, are acquired from the original system, whose function is to remove the impacts of the initial conditions of the reservoir. This process is carried out in the ``offline” fashion (before substituting the target). After that, the machine is incorporated into the system as the digital twins of oscillator $i$. In doing this, the machine receives still coupling signals, $\bm{u}^c_i (t)$, from other oscillators in the system (through the coupling channel) and is also driven by the intrinsic parameter $\omega_i$ (through the parameter-control channel), whereas $\bm{u}^s_i (t)$ is replaced by $\bm{v}(t)$ (in the state channel). From the output vector $\bm{v} = (v_x,v_y)$, we estimate the instant phase of the virtual oscillator, $\theta_i(t) = \arctan(v_x/v_y) + \pi/2$ for $v_x > 0$ and $\theta_i(t) = \arctan(v-x/v-y) + 3\pi/2$ for $v_x < 0$, which is coupled to other oscillators in the system according to Eq.~(\ref{model}). (When more than one oscillators are substituted, the same machine is deployed at different targets.) The substitution is regarded as successful if the collective behavior of the system, namely the synchronization order parameter, is consistent with that of the original system for a sufficiently long period, e.g., several Lyapunov times.

\subsection{Substituting one oscillator}

\begin{figure*}[tbp]
\begin{center}
\includegraphics[width=0.9\linewidth]{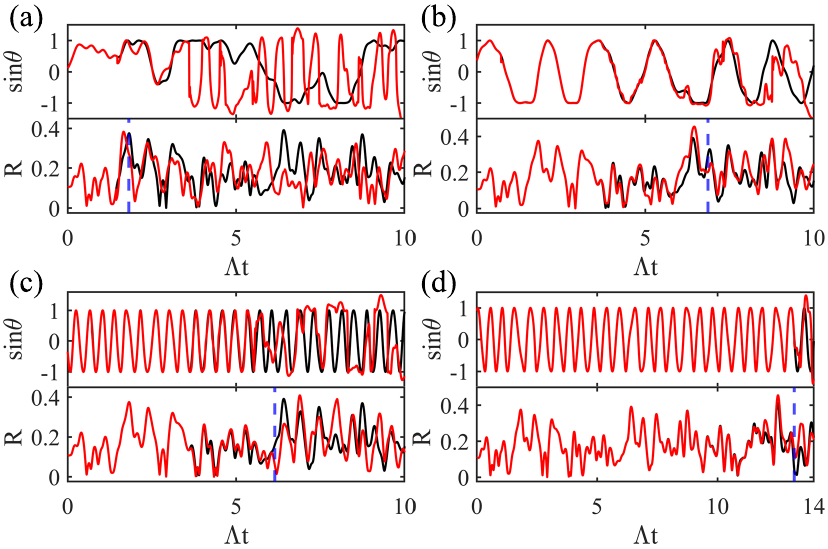}
\centering\caption{Typical results obtained by substituting a single oscillator with the machine. Plotted in each subplot are the time evolutions of the phase of the oscillator (the upper panel) and the synchronization order parameter of the system (the lower panel). (a) The natural frequency of the substituted oscillators is $\omega = -0.044$, which is not included in the sampling set. The oscillator presents irregular motion. The synchronization order parameter can be sustained accurately for about $2$ Lyapunov times. (b) The results for the oscillator of natural frequency $\omega = 0.134$, which is within the sampling set and presents irregular motion. The sustaining period is about $7$ Lyapunov times. (c) The results for the oscillator of natural frequency $\omega = 1.08$, which is not included in the sampling set and presents regular motion. The sustaining period is about $6$ Lyapunov times. (d) The results for the oscillator of natural frequency $\omega = -0.88$, which is included in the sampling set and, in the meanwhile, presents regular motion. The sustaining period is about $13$ Lyapunov times. Black curves are the ground truth. Red curves are the results with oscillator substitution. Vertical dashed lines denote the sustaining horizons.}
\vspace{-0.3cm}
\label{fig2}
\end{center}
\end{figure*}

We check first the performance of adaptable RC in substituting a single oscillator. Plotted in Fig.~\ref{fig2} are the typical results for oscillators that are chosen randomly in the system. In each subplot of  Fig.~\ref{fig2}, the upper panel shows the time evolution of the state of the substituted oscillator, and the lower panel displays the time evolution of the synchronization order parameter of the whole system. The results of the original and machine-substituted systems are represented by the black and red curves, respectively. Shown in Fig.~\ref{fig2}(a) are the results for the oscillator of natural frequency $\omega = -0.044$, which is not included in the sampling set and the time evolution of the phase variable is irregular. We see that when this oscillator is substituted by the machine, the phase state of the oscillator and the synchronization order parameter of the system are sustained accurately for only about $2$ Lyapunov times (defined as the reciprocal of the largest Lyapunov exponent). Shown in Fig.~\ref{fig2}(b) are the results for the oscillator with the natural frequency $\omega = 0.134$, which is one of the sampling oscillators, and the motion of the oscillator is irregular. We see in this case that the state of the oscillator and the synchronization order parameter are sustained accurately for about $7$ Lyapunov times. Shown in Fig.~\ref{fig2}(c) are the results for the oscillator of natural frequency $\omega = 1.08$, which is not included in the sampling set but the phase is evolving with time periodically. In this case, the state and the synchronization order parameter are sustained accurately for about $6$ Lyapunov times. Shown in Fig.~\ref{fig2}(d) are the results for the oscillator of natural frequency $\omega = -0.88$, which is included in the sampling set and the motion of the oscillator is periodic. In this case, the state and the synchronization order parameter are sustained accurately for about $13$ Lyapunov times. Testing results thus show that the performance of the machine is dependent on the properties of the substituted oscillator, including (1) whether the oscillator is within the sampling set (sampling oscillators have longer sustaining periods) and (2) whether the motion of the oscillator is regular (oscillators with regular motions have longer sustaining periods).

\begin{figure}[tbp]
\begin{center}
\includegraphics[width=0.75\linewidth]{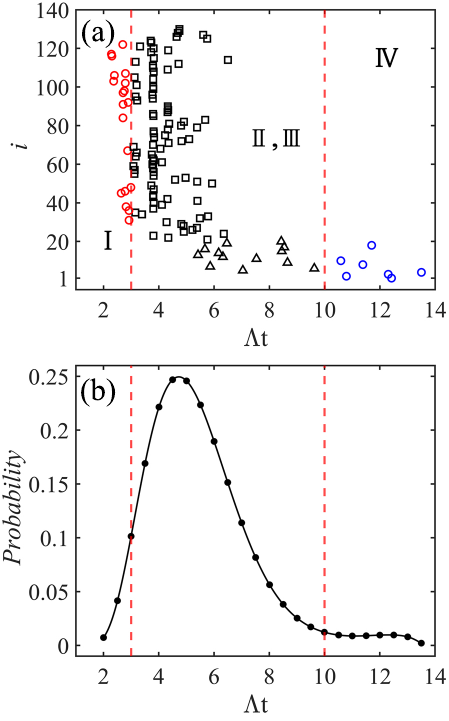}
\centering\caption{A global analysis of the performance of the machine in substituting a single oscillator. (a) The sustaining horizons of the oscillators. Group I (red circles): oscillators that are not included in the sampling set and show irregular motions. Group II (black squares): oscillators that are included in the sampling set but show irregular motions. Group III (black triangles): oscillators that are not included in the sampling set but show regular motions. Group IV (blue circles): oscillators that are not included in the sampling set and show, meanwhile, irregular motions. (b) The distribution of the sustaining horizons. Vertical dashed lines denote the divisions of the oscillators.}
\vspace{-0.5cm}
\label{fig3}
\end{center}
\end{figure}

To evaluate the performance of the machine systematically, we check the sustaining horizon of the system’s collective dynamics for each individual oscillator. Here, the sustaining horizon of an oscillator is defined as the first time the difference between the synchronization parameters of the original and substituted system exceeds the critical value $0.01$ (the results are qualitatively the same when the critical value is slightly changed). The results are presented in Fig.~\ref{fig3}(a). To highlight the dependence of the machine performance on the oscillator properties, we divide the oscillators into four different groups: Group I is composed of oscillators that are not included in the sampling set and present irregular motions [see Fig.~\ref{fig2}(a)]; group II consists of oscillators that are within the sampling set and present irregular motions [see Fig.~\ref{fig2}(b)]; group III contains oscillators which are not included the sampling set but present regular motions [see Fig.~\ref{fig2}(c)]; group IV is made up of oscillators that are included in the sampling set and show, meanwhile, regular motions [see Fig.~\ref{fig2}(d)]. The averaged sustaining horizons of the oscillators in groups I, II, III, and IV are about, respectively, $2$, $6$, $7$, and $10$ Lyapunov times. Clearly, the machine performance is oscillator-dependent: the sustaining duration is long (short) when the substituted oscillator is within (not within) the sampling set and shows regular (irregular) motion, and medium performance is obtained when the oscillator is either not within the sampling set or presenting irregular motion.

For a better characterization of the machine performance, we plot in Fig.~\ref{fig3}(b) the distribution of the sustaining horizons of the oscillators. We see in Fig.~\ref{fig3}(b) that the sustaining horizons are of broad distribution, with the majority of the substitutions distributed within the interval $\Lambda t \in (3, 10)$. The distribution of the sustaining horizons, together with the averaged sustaining horizon over all the oscillators, are the two metrics we adopt for characterizing the overall performance of the machine. For the scenario of single-oscillator substitution studied in Fig.~\ref{fig3}, the averaged sustaining horizon is about $5$ Lyapunov times. In what follows, we are going to demonstrate that the overall performance of the machine can be affected by multiple factors, including the number of substituted oscillators, the number of sampled oscillators, and the uniform coupling strength.

\subsection{Impacts of substitutions, sampling oscillators and coupling strength on machine performance}

\begin{figure}[tbp]
\begin{center}
\includegraphics[width=0.8\linewidth]{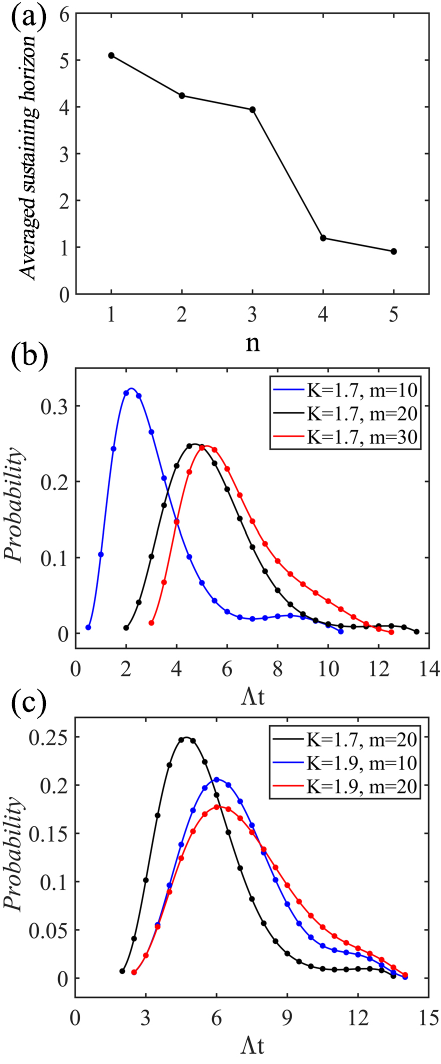}
\centering\caption{The impacts of (a) the number of substituted oscillators, (b) the number of sampling oscillators generating the datasets, and (c) the uniform coupling strength on the performance of the machine. Each result in (a) is averaged over $2000$ realizations for multiple substitutions ($n > 1$). See the context for details.}
\vspace{-0.45cm}
\label{fig4}
\end{center}
\end{figure}

Whereas our interest is focusing on the temporary failure of a single oscillator, the proposed machine-learning scheme can also be applied to situations when multiple oscillators are malfunctioning. In such a case, the same machine is deployed at different oscillators in the system. Due to the chaotic nature of the system dynamics, it is natural to expect that the overall performance will be deteriorated when multiple oscillators are substituted by the machine. To check it out, we utilize the same machine to substitute $n$ oscillators chosen randomly in the system and calculate the sustaining horizon by monitoring the evolution of the synchronization order parameter (as did for single-oscillator substitution). Since the machine performance is dependent on the oscillator properties (as depicted in Fig.~\ref{fig3}), we average the results over a large number of realizations (each realization corresponds to a specific set of substituted oscillators). The results are plotted in Fig.~\ref{fig4}(a). We see that, in agreement with the expectation, the averaged sustaining horizon is decreased as the number of substituted oscillators is increased. In specific, as $n$ is increased from $3$ to $4$, the averaged sustaining horizon is decreased sharply from about $4$ Lyapunov times to about $1$ Lyapunov time. For the results depicted in Fig.~\ref{fig4}(a), we restrict the application of the proposed learning scheme to only situations when a handful of elements in a large-size complex system are temporarily malfunctioning and should be substituted by the machine.

We next investigate the impacts of the sampling oscillators on the overall performance. Specifically, we check how the distribution of the sustaining horizons in the scenario of single-oscillator substitution is affected by the number of sampled oscillators, $m$. Shown in Fig.~\ref{fig4}(b) are the results for $m = 10$ and $30$ sampled oscillators. The new results are compared with the reference results of $m = 20$ shown in Fig.~\ref{fig3}. Oscillators in the sampling set of size $m = 10$ are chosen randomly from the reference set of $m = 20$, and the sampling set of size $m = 30$ is generated by adding $10$ more oscillators (chosen randomly from the system) into the reference set. Two new machines are optimized and trained based on the datasets of the new sampling sets. Figure~\ref{fig4}(b) shows that by increasing (decreasing) the size of the sampling set, the overall performance of the machine is improved (deteriorated). (The averaged sustaining horizons are about $2$ and $6$ for $m = 10$ and $30$, respectively.) The new results are understandable, as the generalization ability of adaptable RC in inferring new dynamics is improved by adopting more sampling states~\cite{KLW:2021,RC:FHW2021,RC:ZH2021,RC:LHB2024}. It should be pointed out that, as more samplings imply an increase in the size of the training dataset and require more computational resources in machine training, we need to strike a balance between the performance and cost in real applications.

We move on to investigate the impacts of the uniform coupling strength on the performance of the machine. As depicted in Fig.~\ref{fig1}(a), with the increase of the uniform coupling strength, the synchronization degree of the oscillators becomes larger and the oscillators tend to move in unison, which favors the learning of the system dynamics in general. For instance, in the extreme case of global synchronization ($K \gg 1$ and $R = 1$), the learning of a single periodic oscillator is enough for inferring the collective dynamics. To demonstrate the beneficial effects of stronger couplings on machine learning, we increase the coupling strength to $K = 1.9$ (by which the synchronization order parameter is about $R = 0.27$ and the largest Lyapunov exponent is about $\Lambda = 0.09$) and evaluate the performance of the new machine trained by the times series of the same set of oscillators studied in Fig.~\ref{fig3} ($m = 20$). The results are plotted in Fig.~\ref{fig4}(c). We see that, compared to the results of $K = 1.7$, the machine performance is clearly improved for $K = 1.9$ (the averaged sustaining horizon is about $8$ Lyapunov times). To illustrate the improved performance under strong couplings further, we plot in Fig.~\ref{fig4}(c) also the results for a reduced sampling set [the set containing $m = 10$ oscillators studied in Fig.~\ref{fig4}(b)]. We see that, despite the reduced sampling set, the overall performance is still improved (the averaged sustaining horizon is about $7$ Lyapunov times), as compared with the results of $K = 1.7$ (the averaged sustaining horizon is about $5$ Lyapunov times). Comparing the results displayed in Figs.~\ref{fig4}(b) and (c), it is also observed that by increasing $m$ from $10$ to $20$, the improvement of the overall performance is more prominent for $K = 1.7$ than for $K = 1.9$. This phenomenon is attributed to the high-degree synchronization of the oscillators under strong couplings, as, in general, the stronger the coherence of the oscillators, the less information given by an additional oscillator about the system dynamics.

\subsection{Sustainability of functional networks}

While the functionality of complex engineering systems such as power grids relies on the precise coordination of the states of their constituent elements, for complex biological systems like neuronal networks, it is often the correlative relationships among the dynamical elements that fundamentally underpin their functionality, namely the functional networks~\cite{FN:BB1995,FN:CR2005,FN:EB2009,FN:KJF2011}. Briefly, functional networks represent the pairwise correlations among coupled dynamical elements within a complex system. It underscores the coordinated dynamics of the system components at the functional level and serves as the dynamical foundation for the functions of a wide range of real-world systems, such as the cognitive processes in the human brain~\cite{FN:KJF2011}. Unlike structural networks, where links represent physical connections, links in functional networks are virtual and defined by correlation criteria. As such, different functional networks can be derived from the same structural network, depending on the definitions of element correlation and the correlation criteria~\cite{FN:ZCS2006,FN:LMH2010,FN:LWJ2015}. Recently, functional networks have also been employed for exploring the working mechanisms of artificial neural networks in machine learning, in which a variety of interesting phenomena have been revealed~\cite{RC:WL2022,RC:LZX2020,RC:WL2022}. Since functional links are determined by the correlations among coupled elements, functional networks effectively characterize the long-term statistical properties of the system dynamics, namely the system’s ``dynamics climate”~\cite{RC:Pathak2017,RC:LZ2018}. Given the profound implications of functional networks for the functionality of complex systems, a question of interest therefore is: When the machine fails to sustain the long-term system evolution accurately, can the functional network still endure?

\begin{figure*}[bpt]
\begin{center}
\includegraphics[width=0.9\linewidth]{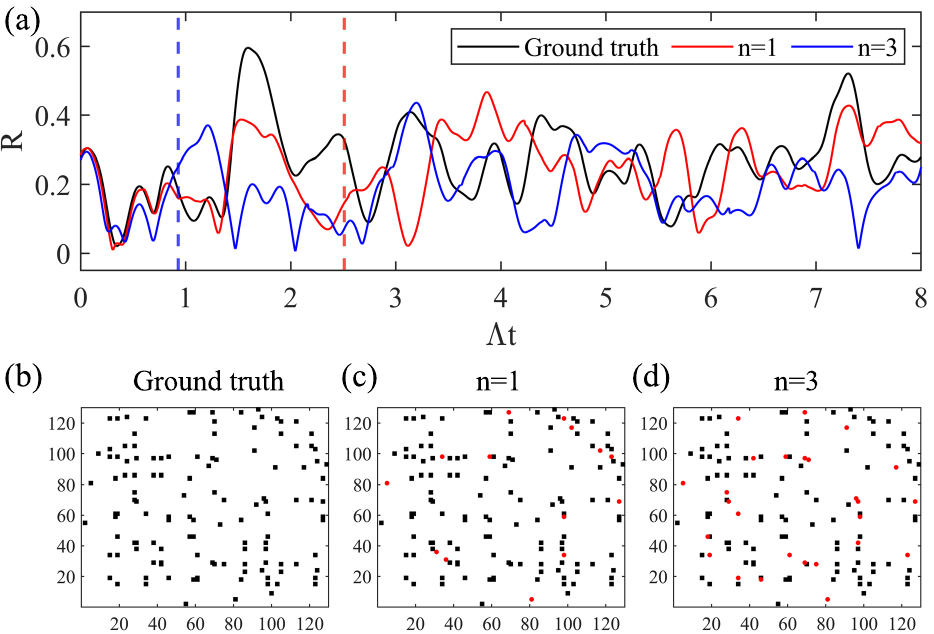}
\caption{Sustainability of functional networks. (a) The time evolution of the synchronization order parameter, $R(t)$. Black curve represents the result of the original system. Red curve is the result for substituting a single oscillator. Blue curve is the result for substituting $n = 3$ oscillators. (b) The functional network of the original system. (c) The functional network of the system with one substitution. About $94\%$ of the links are maintained. (d) The functional network of the system with $n = 3$ substitutions. About $90\%$ of the links are maintained. Red dots in (c) and (d) represent the links that are missed or falsely constructed in the functional networks of the substituted systems.}
\vspace{-0.3cm}
\label{fig5}
\end{center}
\end{figure*}

To investigate, we select the substitution yielding the poorest performance for the results of $K = 1.9$ and $m = 20$ shown in Fig.~\ref{fig4}(c) (the natural frequency of the substituted oscillator is $\omega = -0.609$) and analyze the long-term statistical properties of the system dynamics from the perspective of functional networks. Plotted in Fig.~\ref{fig5}(a) is the time evolution of the synchronization order parameter of the original and substituted systems, which depicts that the system dynamics is sustained accurately for merely about $2.5$ Lyapunov times. The functional networks of the original and substituted systems are presented in Figs.~\ref{fig5}(b) and (c), respectively. Here, the functional networks are constructed based on the Pearson correlation matrix, with the threshold correlation coefficient being chosen as $0.85$ (the results to be reported are qualitatively the same if the threshold is slightly changed). We see that the functional network of the substituted system is in good agreement with that of the original system. Specifically, among the $124$ functional links in the original system, $117$ of them are well maintained in the substituted system, achieving a success rate about $94\%$. We note that this success rate is obtained for the poorest substitution in Fig.~\ref{fig4}(c) ($K = 1.9$ and $m = 20$). For the typical substitutions, the success rates are close to $100\%$. To demonstrate the sustainability of functional networks further, we choose by random two more oscillators in the system and substitute them with the machine too (i.e., $n = 3$ oscillators are substituted in total). The time evolution of the substituted system dynamics is plotted in Fig.~\ref{fig5}(a), which shows that the synchronization order parameter is sustained for less than $1$ Lyapunov time. The functional network of the substituted system is plotted in Fig.~\ref{fig5}(d). In this case, $111$ links are sustained and the success rate is about $90\%$. Results thus show that even though the precise system dynamics is sustained for only a short period, the functional network of the system can be sustained over the long-term system evolution.

\section{Discussions and conclusion}

Though the technique of RC has been applied successfully to the inference of chaos in a variety of low-dimensional systems, its application to spatially extended systems characterized by heterogeneous dynamics and complex coupling structures remains a significant challenge~\cite{RC:Pathak2018,RC:Parlitz2018,RC:ParallelPRL2022}. Different from existing studies that focus on predicting the dynamics of all elements in spatially extended systems, our research concentrates on the sustainability of the collective dynamics of a complex system in the presence of occasional and temporary element failures. Moreover, we consider the general scenario that the system elements are of heterogeneous dynamics, and our primary goal is to train a versatile machine capable of substituting any element in the system while preserving the system’s collective dynamics, namely the synchronization order parameter, over a sufficiently long period. Two features distinguishing the current study from the existing ones are that (1) the machine is of small size and (2) a single machine is able to substitute any element in the system. The former effectively circumvents the problem of ``curse of dimensionality” encountered in the machine learning of spatially extended systems. And the latter significantly reduces the vast computational cost required for the machine learning of heterogeneous complex systems~\cite{RC:Pathak2018,RC:Parlitz2018,RC:ParallelPRL2022}. Though our studies are motivated by the maintenance of the functionality of complex systems suffering from occasional and temporary element failures (e.g., the power grids), the scheme we have proposed might have broad applications, saying, for instance, the virtual neurons in brain-computer interface, the digital twins in multi-agent systems, and the virtual characters in social complex systems. Additionally, our studies provide an alternative approach to the machine learning of spatially extended systems (by proposing the strategy of element substitution), which create a new avenue to the model-free prediction of heterogenous complex systems (by proposing the scheme of versatile RC).

The findings of our present work advance the studies of RC in two aspects. First, the current study extends the application of adaptable RC to spatially extended complex systems. As an effective approach to achieving knowledge transfer in machine learning, the technique of PARC has been widely adopted for inferring new dynamics not included in the training data~\cite{RC:Pathak2018,RC:Parlitz2018,RC:ParallelPRL2022}. However, the existing studies are restricted to either low-dimensional systems or complex systems featured by homogeneous dynamics (i.e., the Lorenz-96 climate model), and it remains unclear whether the technique can be generalized to spatially extended systems characterized by heterogeneous dynamics. Our studies show that, by incorporating the coupling channel into the standard PARC architecture, the machine is able to replicate precisely not only the dynamics of the sampled elements, but also the dynamics of new elements not included in the sampling set. Second, the new learning scheme proposed in the present work paves the way for inferring the dynamics of heterogeneous complex systems. Though the scheme of parallel RC has been proven efficient for learning spatially extended systems (i.e., the Kuramoto-Sivashinsky model), its application to complex systems with heterogeneous dynamics remains a great challenge. In Ref.~\cite{RC:ParallelPRL2022}, the authors generalized the parallel RC scheme and utilized it to forecast the dynamics of networked non-identical phase oscillators (a generalized Koramoto model). There, to cope with the heterogeneous dynamics of the oscillators, each oscillator is emulated by a unique machine, with the machines being optimized and trained individually. Our research advances the studies in Ref.~\cite{RC:ParallelPRL2022} by demonstrating that the diverse dynamics of the oscillators can be replicated by a single machine. The all-in-one feature of the new machine not only reduces the computational cost for machine training and optimization, but also brings great flexibility and convenience in real applications, especially in scenarios when a few elements experience temporary failures in a large-size system and virtual substitutions (e.g., digital twins) are adopted for sustaining the system functionality.

A few remarks regarding the limitations and open questions of the proposed learning scheme are as follows. First, the classical Kuramoto model we have adopted is special in that the oscillators, though of different natural frequencies, are globally coupled. While the proposed scheme is anticipated to be applicable also to complex networks with homogeneous structures (e.g., the network model investigated in Ref.~\cite{RC:ParallelPRL2022}), its application to complex networks with heterogeneous structures is still an open question. Second, in substituting an element with the machine, we need to know in advance the intrinsic parameter of the element and, in addition, the set of neighboring oscillators connected to it. These limitations are rooted in the parameter-aware and coupling-guided features of the machine, which can not be avoided in the current learning scheme. Third, compared to the conventional scheme of parallel RC, where each element is emulated by a unique machine, the new learning scheme excels in flexibility (i.e., a single machine is able to substitute any element) but lags in performance. By the conventional parallel scheme, the synchronization order parameter can be sustained accurately for approximately several Lyapunov times~\cite{RC:ParallelPRL2022}; for the new learning scheme proposed in our current study, the sustaining horizon is merely about $1$ Lyapunov time when $n = 5$ oscillators are substituted, as depicted in Fig.~\ref{fig4}(a). This characteristic makes the new learning scheme more suitable for element substitution than for dynamics inference.

To summarize, inspired by the temporary and sporadic failures of elements in complex dynamical systems, we have proposed a new machine-learning scheme in which a single machine is able to substitute any element while preserving the collective dynamics of the large-size complex system for a certain period. The validity and feasibility of the new learning scheme have been justified and demonstrated by substituting oscillators in the classical Kuramoto model, and the performance of the machine has been systematically evaluated. It was revealed that the machine performance is dependent on multiple factors, including the characteristics of the substituted element (such as whether it belongs to the sampling set and exhibits irregular motion), the training cost (measured by the number of sampled oscillators), the number of substitutions, and the uniform coupling strength. Additionally, it was discovered that although the machine fails to sustain the system dynamics precisely over a long period, the statistical properties of the system dynamics, namely the functional networks, are well maintained in the long-term evolution. Our studies provide an alternative approach to the machine learning of spatially extended complex systems characterized by heterogeneous dynamics, and the new learning scheme might have implications for the maintenance and operation of a variety of real-world complex dynamical systems experiencing temporary and sporadic element failures.

The source codes and data that support the findings of this study are available from the corresponding author upon reasonable request.

\begin{acknowledgments}
This work was supported by the National Natural Science Foundation of China (NNSFC) under Grant Nos.~12275165 and 12105165. X.G.W. was also supported by the Fundamental Research Funds for the Central Universities under Grant No.~GK202202003.
\end{acknowledgments}


\begin{thebibliography}{99}
\bibitem{RC:Maass2002} W. Maass, T. Natschlager, and H. Markram, Real-time computing without stable states: A new framework for neural computation based on perturbations, Neural Comput. {\bf 14}, 2531 (2002).

\bibitem{RC:Jaeger} H. Jaeger and H. Haas, Harnessing nonlinearity: Predicting chaotic systems and saving energy in wireless communication, Science {\bf 304}, 78 (2004).

\bibitem{RC:Pathak2017} J. Pathak, Z. Lu, B. Hunt, M. Girvan, and E. Ott, Using machine learning to replicate chaotic attractors and calculate Lyapunov exponents from data, Chaos {\bf 27}, 121102 (2017).

\bibitem{RC:LZ2018} Z. Lu, B. R. Hunt, and E. Ott, Attractor reconstruction by machine learning, Chaos {\bf 28}, 061104 (2018).

\bibitem{RC:Fan} H. Fan, J. Jiang, C. Zhang, X. G. Wang, and Y.-C. Lai, Long-term prediction of chaotic systems with machine learning, Phys. Rev. Res. {\bf 2}, 012080(R) (2020).

\bibitem{RC:digtwin} L.-W. Kong, Y. Weng, B. Glaz, M. Haile, and Y.-C. Lai, Reservoir computing as digital twins for nonlinear dynamical systems, Chaos {\bf 33}, 033111 (2023). 

\bibitem{RC:DY2024} Y. Du, Q. Li, H. Fan, M. Zhan, J. Xiao, and X. G. Wang, Inferring attracting basins of power system with machine learning, Phys. Rev. Res. {\bf 6}, 013181 (2024).

\bibitem{RC:NC2024perspective} M. Yan, C. Huang, P. Bienstman, P. Tino, W. Lin, and J. Sun, Emerging opportunities and challenges for the future of reservoir computing, Nat. Commun. {\bf 15}, 2056 (2024).

\bibitem{RC:adaptiveRCLYC} S. Panahi and Y.-C. Lai, Adaptable reservoir computing: A paradigm for model-free data-driven prediction of critical transitions in nonlinear dynamical systems, Chaos {\bf 34}, 051501 (2024).

\bibitem{RC:lukosevicius2009} M. Lukosevicius and H. Jaeger, Reservoir computing approaches to recurrent neural network training, Comput. Sci. Rev. {\bf 3}, 127 (2009).

\bibitem{RC:Tanaka2019} G. Tanaka, T. Yamane, J. B. Heroux, R. Nakane, N. Kanazawa, S. Takeda, H. Numata, D. Nakano, and A. Hirose, Recent advances in physical reservoir computing: A review, Neural Networks {\bf 115}, 100 (2019).

\bibitem{RC:Book} K. Nakajima and I. Fischer, {\it Reservoir Computing: Theory, Physical Implementations, and Applications} (Springer, Singapore, 2021).

\bibitem{RC:CK2020} C. Klos, Y. F. K. Kossio, S. Goedeke, A. Gilra, and R.-M. Memmesheimer, Dynamical learning of dynamics, Phys. Rev. Lett. {\bf 125}, 088103 (2020).

\bibitem{RC:Guo2021} Y. L. Guo, H. Zhang, L. Wang, H. W. Fan, J. H. Xiao, and X. G. Wang, Transfer learning of chaotic systems, Chaos {\bf 31}, 011104 (2021).

\bibitem{KLW:2021} L.-W. Kong, H.-W. Fan, C. Grebogi, and Y.-C. Lai, Machine learning prediction of critical transition and system collapse, Phys. Rev. Res. \textbf{3}, 013090 (2021).

\bibitem{RC:Kim2021} J. Z. Kim, Z. Lu, E. Nozari, G. J. Pappas, and D. S. Bassett, Teaching recurrent neural networks to infer global temporal structure from local examples, Nat. Mach. Intell. {\bf 3}, 316 (2021).

\bibitem{RC:FHW2021} H. Fan, L.-W. Kong, Y.-C. Lai, and X. G. Wang, Anticipating synchronization with machine learning, Phys. Rev. Res. {\bf 3}, 023237 (2021).
%
\bibitem{RC:XR2021} R. Xiao, L.-W. Kong, Z.-K. Sun, and Y.-C. Lai, Predicting amplitude death with machine learning, Phys. Rev. E {\bf 104}, 014205 (2021). 

\bibitem{RC:ZH2021} H. Zhang, H. Fan, L. Wang, and X. G. Wang, Learning Hamiltonian dynamics with reservoir computing, Phys. Rev. E {\bf 104}, 024205 (2021).

\bibitem{HWFan2022} H. Fan, L. Wang, Y. Du, Y. F. Wang, J. H. Xiao, and X. G. Wang, Learning the dynamics of coupled oscillators from transients, Phys. Rev. Res. \textbf{4}, 013137 (2022).

\bibitem{RC:multistability2022} M. Roy, S. Mandal, C. Hens, A. Prasad, N.V. Kuznetsov, and M. D. Shrimali, Model-free prediction of multistability using echo state network, Chaos {\bf 32}, 101104 (2022).

\bibitem{PARC:Jalan2024} D. Sisodia and S. Jalan, Dynamical analysis of a parameter-aware reservoir computer, Phys. Rev. E {\bf 110}, 034211 (2024).

\bibitem{RC:LHB2024} H. Luo, Y. Du, H. Fan, X. Wang, J. Guo, and X. G. Wang, Reconstructing bifurcation diagrams of chaotic circuits with reservoir computing, Phys. Rev. E {\bf 109}, 024210 (2024).

\bibitem{KLW2024} L.-W. Kong, G. A. Brewer, and Y.-C. Lai, Reservoir computing based associative memory and itinerancy for complex dynamical attractors, Nat. Commun. {\bf 15}, 4840 (2024).

\bibitem{YD:MFRC} Y. Du, H. Luo, J. Guo, J. Xiao, Y. Yu, and X. G. Wang, Multifunctional reservoir computering, Phys. Rev. E {\bf 111}, 035303 (2025).

\bibitem{COD} R. Bellman, Dynamic programming, Science {\bf 153} 34 (1966).

\bibitem{RC:Pathak2018} J. Pathak, B. Hunt, M. Girvan, Z. Lu, and E. Ott, Model-free prediction of large spatiotemporally chaotic systems from data: A reservoir computing approach, Phys. Rev. Lett. {\bf 120}, 024102 (2018).

\bibitem{RC:Parlitz2018} R. S. Zimmermann and U. Parlitz, Observing spatio-temporal dynamics of excitable media using reservoir computing, Chaos {\bf 28}, 043118 (2018).

\bibitem{RC:ParallelPRL2022} K. Srinivasan, N. Coble, J. Hamlin, T. Antonsen, E. Ott, and M. Girvan, Parallel machine learning for forecasting the dynamics of complex networks, Phys. Rev. Lett. {\bf 128}, 164101 (2022).

\bibitem{RC:AtmosphereForecastOtt2020} I. Szunyogh, T. Arcomano, J. Pathak, A. Wikner, B. Hunt, and E. Ott, A machine learning-based global atmospheric forecast model, Geophys. Res. Lett. {\bf 47}, e2020GL087776 (2020).

\bibitem{Kuramoto:Book}Y. Kuramoto, {\it Chemical Oscillations, Waves, and Turbulence} (Springer, Berlin, 1984).

\bibitem{FN:BB1995} B. Biswal, F. Z.  Yetkin, V. M. Haughton, and J. S. Hyde, Functional connectivity in the motor cortex of resting human brain using echo-planar MRI, Magnetic Resonance in Medicine {\bf 34}, 537 (1995).

\bibitem{FN:CR2005} C. Ranganath, A. Heller, M. X. Cohen, C. J. Brozinsky, and J. Rissman, Functional connectivity with the hippocampus during successful memory formation, Hippocampus {\bf 15}, 997 (2005). 

\bibitem{FN:EB2009} E. Bullmore and O. Sporns, Complex brain networks: Graph theoretical analysis of structural and functional systems, Nat. Rev. Neurosci. {\bf 10}, 186 (2009).

\bibitem{FN:KJF2011} K. J. Friston, Functional and effective connectivity: A review, Brain Connect. {\bf 1}, 13 (2011).

\bibitem{FN:ZCS2006} C. S. Zhou, L. Zemanova, G. Zamora, C. C. Hilgetag, and J. Kurths, Hierarchical organization unveiled by functional connectivity in complex brain networks, Phys. Rev. Lett. {\bf 97}, 238103 (2006).

\bibitem{FN:LMH2010} M. Li, X. G. Wang, and C. H. Lai, Evolution of functional subnetworks in complex systems, Chaos {\bf 20}, 045114 (2010).

\bibitem{FN:LWJ2015} W. Lin, Y. Wang, H. Ying, Y.-C. Lai, and X. G. Wang, Consistency between functional and structural networks of coupled nonlinear oscillators, Phys. Rev. E {\bf 92}, 012912 (2015).

%
\bibitem{RC:LZX2020} Z. Lu and D. S. Bassett, Invertible generalized synchronization: A putative mechanism for implicit learning in neural systems, Chaos {\bf 30}, 063133 (2020).

\bibitem{RC:WL2022} L. Wang, H. Fan, J. Xiao, Y. Lan, and X. G. Wang, Criticality in reservoir computer of coupled phase oscillators, Phys. Rev. E {\bf 105}, L052201 (2022).
%
\end{thebibliography}
\end{document}